# Understanding And Affecting Student Reasoning About Sound Waves

Michael C. Wittmann
*Department of Physics and Astronomy, University of Maine*
*5709 Bennett Hall*
*Orono, ME 04469-5709*
wittmann@umit.maine.edu, tel: 207 – 581 – 1237

Richard N. Steinberg
*Department of Physics, City College of New York*

Edward F. Redish
*Department of Physics, University of Maryland*

## Abstract

Student learning of sound waves can be helped through the creation of group-learning classroom materials whose development and design rely on explicit investigations into student understanding. We describe reasoning in terms of sets of <u>resources</u>, i.e. grouped building blocks of thinking that are commonly used in many different settings. Students in our university physics classes often used sets of resources that were different from the ones we wish them to use. By designing curriculum materials that ask students to think about the physics from a different view, we bring about improvement in student understanding of sound waves. Our curriculum modifications are specific to our own classes, but our description of student learning is more generally useful for teachers. We describe how students can use multiple sets of resources in their thinking, and raise questions that should be considered by both instructors and researchers.

## Introduction

Research-based curriculum development programs have been shown to be effective in helping students gain a conceptual understanding of many specific topics in physics (McDermott & Redish, 1999). In describing the success of many of these materials, a special focus is rightly given to the specific details of investigations of students at the developing school. Instructors wishing to implement research-based materials at other locations may not have similar populations, though, and may require additional information to help them improve their own teaching. These readers may wish for a more general discussion of how specific methods of instruction are chosen in relation to the specific issues raised in the research. For such an audience, the theoretical approaches that guide curriculum development may be as important as the specific tasks built into the classroom materials and the specific research data on which they are based. Two elements, a detailed picture of student understanding of the physics of sound waves and a detailed model of learning, are necessary to inform these instructors how to approach students in the classroom and recognize student reasoning not specifically described in this paper. We discuss this double approach in the context of student learning of the wave physics of sound.

We have reported previously on student reasoning about propagation and superposition (Wittmann, Steinberg, & Redish, 1999) and the mathematical description of waves (Steinberg, Wittmann, & Redish, 1997). In all contexts, we find (Wittmann, 2001) that many students approach wave physics using reasoning that is problematic and unproductive in its focus on object-like properties rather than properties related to propagation of cascading chains of events. In this paper, we show that students use a similar problematic approach when discussing sound waves.

Though the physics of sound is frequently taught in the introductory sequence of university physics, student understanding of this difficult topic has been studied only sparingly. The results of Cedric Linder and collaborators (Linder & Erickson, 1989; Linder, 1992, 1993) show that many students are unable to interpret common interpretations of sound waves, are unable to account for the underlying wave model of sound, or have no coherent model of the physics. These results are consistent with physics education research findings in many other areas of introductory (McDermott, 1991; Beichner, 1994) and advanced physics (Steinberg, Oberem, & McDermott, 1996). We build on their work and focus specifically on how a model of learning plays a role in understanding their reasoning.

Our research was carried out at the University of Maryland with engineering students in the second semester of a three-semester introductory physics sequence. Each week, students had three hours of lecture, three hours of laboratory (some of which dealt with standing mechanical waves on a string, but none of which dealt with sound waves), and one hour of discussion section. These discussion sections were either traditional problem-solving sessions or University of Washington-style Tutorials (McDermott, Shaffer, & Washington, 1998). In Tutorials, students work in groups on activities designed through the process of physics education research to help them come to a better understanding of difficult physics concepts (McDermott, Shaffer, & Somers, 1994; Redish, Saul, & Steinberg, 1997; Steinberg et al., 1997; Wosilait, Heron, & McDermott, 1998).

In this paper, we describe how one Tutorial, occupying only an hour of a student's weekly curriculum, can have a measurable and positive effect on student learning in the context of sound waves. The Tutorial was developed based on a fine-grained view of student reasoning in which basic reasoning elements that are intuitively appropriate in many situations are applied incorrectly in the context of waves (this model is described in more detail below).

In carrying out this research, we were interested in three topics. These can be summarized as:

- How do students make sense of sound waves in a context they are unaccustomed to describing formally?
- How can we create an instructional context that best helps them choose an appropriate model of sound waves, and how is this instructional context connected to the reasoning skills they bring to the classroom?
- How can the description of student reasoning and the model of the physics help instructors who observe inconsistent student reasoning in the classroom?

To describe how students think about sound waves requires two elements. We must describe the general way in which we understand student sense-making in our classroom. We must also describe specifically how students make sense of sound waves in particular.

## Model Of Student Learning

A specific model of student reasoning guides our discussion. In this model, we analyse student thinking in terms of the often reasonable intuitive physics that they bring to the classroom (Hammer, 2000). In the context of wave physics, we find that students often reason by focusing solely on object-like properties of the system they consider (Wittmann, 2001). This focus guides student reasoning in unfamiliar settings when students associate various visual and mathematical cues with object and not wave properties. Many students state that wave propagation speed depends on the force used to create a wave, superposition occurs only at a single point, and waves collide when meeting each other (or reflecting off surfaces). They are using reasoning resources that are appropriate, respectively, to objects and their motion, their centre of mass, and their interactions (e.g. collisions). In this section of the paper, we introduce the language of learning theory necessary to help us interpret these results. We briefly describe one model of student thinking that helps us make sense of our data (diSessa & Sherin, 1998, diSessa, 1993). These ideas build on previously published descriptions (Wittmann et al., 1999; Wittmann, 2001) and are strongly influenced by Hammer (Hammer, 2000).

Our goal in this section is to show that describing the physics from the point of view of how we understand and think about the world around us is relevant and important to how we approach students in our classes. Our goal is not to illustrate a complete theoretical structure but rather to extract a few general guiding elements and principles from attempts to build more complete theories that can help instructors better understand how to implement and understand the workings of reform curricula. Thus, we do not discuss in detail the relation of our results to more specific theories such as coordination classes (Wittmann, 2002, diSessa & Sherin, 1998) or ontologies (Reiner et al., 2000).

### Thinking based on resources

Consider a situation where you see, out of the corner of your eye, an object dropping toward you from above. What do you do? Jump sideways? Catch it? How does your answer change if you recognize a leaf? or a rock? Your choice might depend on a variety of skills you have. For example, you can estimate its location and motion, as well as its size, or its weight, even after a brief glance. These are all skills which are familiar when thinking about objects. An object's weight is important when it is lifted or pushed. An object's size is important when we estimate its location in the distance while driving. An object's location, in this case its path through the air, is important when throwing balls or looking at the movement of cars on the highway. We have skills for determining and reasoning about each of these qualities of an object. (For example, catch the leaf but avoid the rock.)

Similarly, we have skills for simplifying our reasoning about the world around us and choosing how to interpret a situation. We often choose a point to represent the whole, both in physics and in our daily lives. In television sports broadcasts of football or basketball, commentators use screen pens to represent people with a dot, their motion with a line. (Coaches might use X's and O's.) The representation seems natural and is easily interpreted. Similarly, the sentence, 'a million faces were turned to the screen as the shuttle disaster occurred' indicates far more than faces watching television. The face is used to represent the entire body and attention of a person. Similarly, a nation's president or monarch represents the whole country, the CEO represents the much larger company, and so on. In each case, a single point or element of the larger system is used to simplify thinking about a much larger and much more complicated body. Usually, this body is thought of as an object. (Consider that even certain multinational corporations are referred to as 'monolithic', like a monument.)

In physics, we also often describe extended objects using single points. In introductory physics, we tacitly rely on this, discussing the motion of objects for the first few weeks of the course before we even raise the more complex issue of centre of mass. For the traditional cannonball problem, this might be easily interpretable and understandable, but for a clown shot from a circus cannon, the issue is more difficult. The person might have flailing arms, might be rotating, and might seem anything but point-like. Still, we use the centre of mass to describe a parabolic path through the air. When solving problems in Newtonian dynamics, we often use free-body diagrams in which we abstract a large body to a single point and then discuss the forces acting on this point. The details of the shape of the object are often considered irrelevant to the analysis being done. (This situation obviously changes when a torque problem is done. Moving beyond a single point when describing an entire object is something students often find difficult. For more detail, see Ortiz, Heron, Shaffer, & McDermott, 2000.)

Using a single point to represent an entire body is one example of the many methods that we have to simplify the way we interpret and reason about the world around us. Furthermore, this example shows how we often use basic abstractions in many different situations.

Many different types of such elemental building blocks have been described in the literature. These include phenomenological primitives (diSessa, 1993), facets of knowledge (Minstrell, 1992), intuitive rules (Tirosh, Staby, & Cohen, 1998), and reasoning resources (Hammer, 2000). Each of these is a theoretically distinct type of resource, either conceptual, representational, or perceptual, or some combination thereof. It is not the purpose of this paper to describe them in detail nor to discuss theoretical differences in the terminology. Instead, we will use a general description of student learning based on what we will generally call resources to describe the basic pieces of reasoning that students bring to the classroom. These fine-grain elements of reasoning help us make sense of how student understanding changes due to instruction.

### Sets of resources

Reasoning in physics rarely involves only individual resources. Combining resources into connected sets helps us simplify our reasoning about otherwise complicated phenomena.

Consider the resources we use for thinking about objects. These include perceptual tools that help us estimate size and weight when simply observing an object (e.g. why react differently when we see a leaf or a rock dropping toward us?). We are able to describe the motion of objects as they physically move from one location to the next. We also have conceptual and phenomenological resources (such as bouncing or blocking (diSessa, 1993)) to help interpret interactions between objects (e.g. the truck hitting the car will most likely cause far greater damage than it will receive). We are able to combine these resources into sets that help us quickly interpret a situation and react to it (e.g. while driving a car, do we run over a seemingly empty plastic grocery bag or not, and how does that choice depend on the bag's motion?).

Another set of resources that we can describe involves events. By an event, we mean something described by the sentence 'something happened'. For example, a person standing up and sitting down is an event (or possibly two), while a person simply sitting still is not ('nothing happened'). Note that we purposefully use the colloquial meaning of the term 'event' rather than the rigorous one from, for example, relativity theory.

With events, we also have a set of grouped resources to interpret the world around us. These let us quickly and easily interpret events without necessarily conscious awareness. Specifically, we have resources (perceptual, conceptual, phenomenological, and so on) for thinking about a series of events like a wave in a football stadium,[1] or playing the game 'telephone' as a child. We have reasoning resources to help us understand issues such as the speed of the movement of the event, the effect of the event on its surroundings, and the types of events we see. These resources are particularly useful when trying to explain identical things that happen at different locations (e.g. assuming identical causes for identical events). Note that the same resources may play a role in describing both objects and events. Collisions (things get in each other's way when moving) obviously describe how objects interact, but a series of collisions (imagine a car pile-up on the highway) can be thought of as a causally linked propagating event.

Another important consideration is that when thinking about a series of events, we may be less likely to think of objects. For example, the sports stadium wave is more fascinating (and dominates our perceptions more) than the motion of an individual in the stadium (until we ourselves stand up and sit down, with a loud cheer). This illustrates how thinking about a phenomenon from one point of view (the wave motion) may hamper us from thinking about it from another point of view (the motion of an individual) except in specific circumstances. A clear description of the process would require both views, however. Often, a complete understanding of a situation requires that one think about the objects that play a role when describing events. Similarly, we must often describe the events that objects are subject to.

What often distinguishes novices from experts in their use of resources is the level of familiarity with the different sets of resources that are applicable in a specific setting. In addition, experts have often 'compiled' their knowledge into ideas and concepts that are easily called up and employed in a way that novices are unable to do. (Consider a novice and an expert driver when an object is lying in the road, and who makes quicker choices.) When observing students reasoning about unfamiliar topics, we expect them to use not only inappropriate resources at times, but to be guided by sets of resources that do not match the situation. In this paper, we describe how students often use an inappropriate set of resources to guide their reasoning about sound waves, and how properly designed curriculum can help students come to a better understanding of the physics.

### *Application of model of learning to sound waves*

Of particular interest to this paper is the issue of propagation. Both object propagation (e.g. a thrown ball) and wave propagation (e.g. a sound wave) describe movement from one point to another. A propagating object physically moves from one location to the next. A propagating (mechanical) wave is the motion of a disturbance (e.g. in a sound wave, a displacement from density or pressure equilibrium) from one point to another. These two types of motion have a fundamentally different nature.

When describing the sports stadium wave, both the motion of an individual in the stadium and the motion of the wave play a role in describing movement of something from one point to another. One talks about 'the wave' moving around the stadium as if it were an object physically moving from one location to another. One describes its location and its size as one would describe an object, not an event. (What is the size of an event?) Not until one properly discerns the motion of each person in the wave is it clear that this is a propagating event of people standing up (when those next to them rise) and sitting down (when those next to them sit). With a sports stadium wave, the switch from object-like wave to propagating event is relatively easy. The motions are easily discernible due to their perpendicular directions. More importantly, one can easily imagine oneself participating, allowing for the quick shift between the personal motion one can imagine and the wave motion one can observe.

To think about the different types of motion in the context of sound waves requires focusing on different levels of a system. A proper understanding of sound waves would include applying the correct resources in the correct fashion. Sound waves differ from a sports stadium wave in that the motion of the medium is co-linear with the direction of propagation. In addition, the medium is not easily visible, making a discussion of the distinction between motion of the medium and motion of the wave difficult. Perceptions of sound are auditory and not visual. What is moving must be indirectly inferred. Knowing what is moving becomes difficult when there are no perceptual tools to help.

When we observe students answering questions on the topic of sound wave propagation, we expect that, for the reasons suggested above, they will have problems applying resources appropriate for understanding propagating events rather than propagating objects. In the next section of the paper, we describe such a study.

## Investigating Student Understanding

To investigate how students distinguish between the motion of a wave and the medium through which it travels, we originally posed two different types of questions about sound waves (see figure 1). Some students had to describe the motion of a dust particle sitting motionlessly in front of a previously silent loudspeaker after the speaker had been turned on (figure 1(a)). Others described the motion of a candle flame placed in front of a loudspeaker (figure 1(b)). These questions were administered in both interview and written format before any instruction on waves, during instruction, and after all instruction on waves. Students answered these questions in a variety of settings. More than 25 were interviewed, where they were asked the question in a one-on-one setting that allowed for extended discussion and follow-up on the part of the interviewer. Over 200 answered the questions in a pretest, which is an ungraded conceptual quiz that accompanies Tutorials. (These will be described below.) Finally, 137 students answered these questions in a diagnostic test that was administered before all instruction and 6 weeks after students had completed all instruction on waves. Some students answered both questions, as will be described below.

 

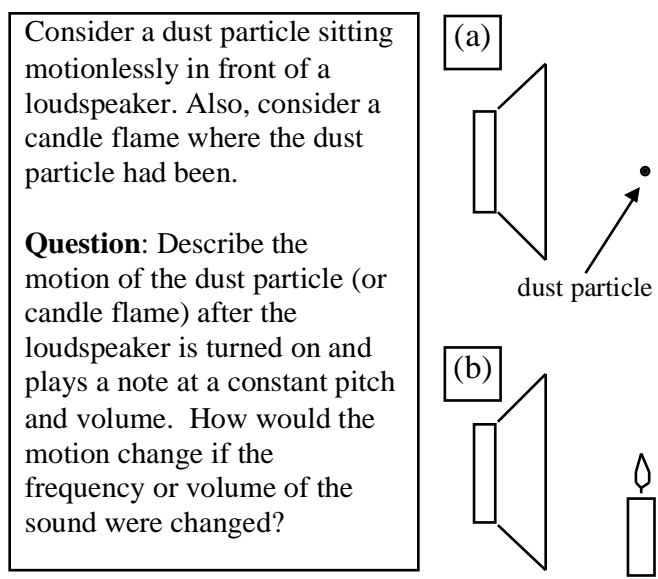

Consider a dust particle sitting motionlessly in front of a loudspeaker. Also, consider a candle flame where the dust particle had been.

**Question**: Describe the motion of the dust particle (or candle flame) after the loudspeaker is turned on and plays a note at a constant pitch and volume. How would the motion change if the frequency or volume of the sound were changed?

(a)

dust particle

(b)

**Figure 1**
Two different situations in which the sound wave question was asked, (a) the dust particle sound wave question, (b) the candle flame sound wave question. In interviews, students were not given a diagram, but had a loudspeaker and a candle and were asked to imagine the dust particle or the candle flame. In pretests and examination questions, students had a variety of diagrams, all equivalent to the ones shown.

The physics of the questions in figure 1 merits discussion. The dust particle, we told the students in interviews, is floating motionlessly in a room with no wind (i.e. there are no outside air currents). This is plausible, when considering that buoyancy can support a dust particle of the right density at the desired height. The lack of air currents is not plausible when considering the candle flame because the heat from the candle causes convection currents. In a simple model of the situation, these currents only occur in the near vicinity of the candle, and, in steady air, add little to the effective size (i.e. width) of the candle. For both systems and at the appropriate size and time scale, we can assume that the medium through which the sound waves travel horizontally is motionless except for the motion from equilibrium caused by the sound waves themselves. In all our interviews, students viewed the situations similarly, and relevant questions dealing with the more detailed physics were not raised.[2] Based on our choice of dust particle size and candle flame size, we can treat them as points in space which move in response to the motion of the medium in which they are embedded. Students, without ever mentioning the detailed physics discussed above, similarly treated the two as point particles.

We expected students to point out that the dust particle and the candle flame would oscillate longitudinally from side to side due to the motion of the air. Rather than dealing with more nuanced and detailed descriptions of the physics, related, for example, to the effect of a flame on a propagating pressure wave, students focused on other aspects of the physics of sound which showed that they had a different model of what was happening. Students regularly treated the waves as objects that are capable of pushing things along in the direction of their motion. This analysis is fundamentally different from one related to the description of propagating events.

### *Discussion of student reasoning*

We found that the basic types of student responses did not change during the course of the semester, but the frequency with which they occurred did change depending primarily on the type of instruction on waves that students had. The understanding of student reasoning that we developed as a result of the interviews was therefore productive in describing what we observed of student thinking at all times during instruction.

We give extended excerpts from two interviews to show the variety of student thinking. In the interviews, we find that students think both in terms of objects and a series of events, yet they tend to focus on the object-like behaviour of the system. We found that most students had great difficulty distinguishing between the propagation of the sound wave and the motion of the medium through which it travels. When viewed from the point of view of an expert, it appears as if the students focus on the compression phase of a sound wave's passage through a medium, but ignore the rarefaction phase. However, we claim that student responses indicate that their reasoning cannot easily be described from an expert physics point of view, since students did not reason using the wave properties of the system.

After discussing two students in detail, we discuss student responses to written questions and discuss the frequency of different types of responses.

## Sound waves kicking the medium in their path

Our first set of interviews was with six students who had completed traditional instruction on waves. One student, Alex, presented reasoning representative of 4 of the 6 students in these interviews. In response to the question in Figure 1, he stated,[3]

> Alex: [The dust particle] would move away from the speaker, pushed by the wave, pushed by the sound wave . I mean, sound waves spread through the air, which means the air is actually moving, so the dust particle should be moving with that air which is spreading away from the speaker.
>
> Interviewer: Okay, so the air moves away --
>
> A: It should carry the dust particle with it.
>
> I: How does [the air] move to carry the dust particle with it?
>
> A: Should push it, I mean, how else is it going to move it? If you look at it, if the particle is here, and this first compression part of the wave hits it, it should move it through, and carry [the dust particle] with it. [See figure 2 for a recreation of the sketch he drew while speaking.]

Alex describes the leading edge of the wave as exerting a (constant) force on the dust particle. Note that he then changes his description, and explicitly states that the wave would 'hit' the dust particle (like one object might hit another).

When asked how his response would change if the original experiment were repeated with the speaker playing a note of twice the frequency, Alex stated:

> A: That would just change the rate at which the particle is moving. . . The wave speed should be, it should double, too. . . Yeah, speed should increase.
>
> I: How did you come to that answer?
>
> A: I was thinking that the frequency of the wave, a normal wave, shows us how many cycles per some period of time we have. . . You can have twice as many cycles here in the same period of time. . .
>
> I: And what effect does it have to go through one cycle versus to go through two cycles?

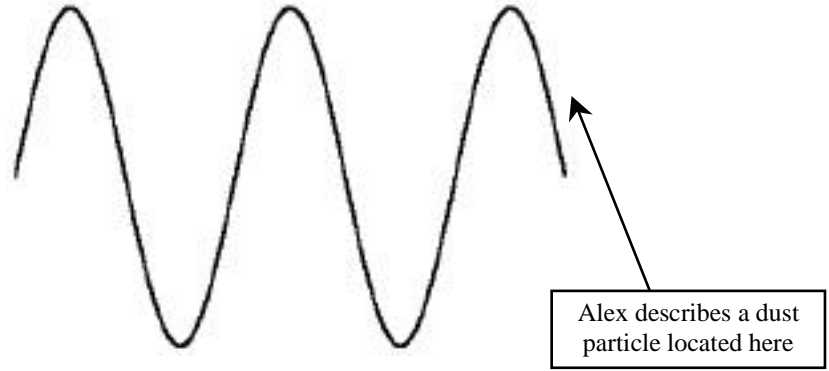

<div style="text-align:center">Alex describes a dust<br/>particle located here</div>

**Figure 2**

Alex's sketch of the sinusoidal sound wave exerting a compression force on the dust particle at its leading edge. The dust particle is "carried along" by the leading edge of the sine curve describing the sound wave.

A: If it goes through one cycle of the compression wave like this, then the first wave should hit it . . . And the second wave which has frequency which is twice as big should hit it twice by then, which should make it go faster.

Alex clearly describes the effect of the change in frequency of the particle: the particle is hit more often by the sound wave. Note that Alex is thinking both in terms of objects (the hitting of one solid object by something, seemingly another solid object) and a series of events (the compression wave hitting the particle repeatedly, and specifically more often in the same amount of time). The following questions were asked to clarify what he meant by the term 'hit'.

I: So each compression wave has the effect of kicking the particle forward?

A: Yeah.

I: So when you've been kicked twice, you're moving twice as fast?

A: Basically, yeah. Right, because the force -- If you have a box, and you apply a force once, the acceleration is - force equals mass times acceleration, you can find the acceleration. [see figure 3 for a recreation of the sketch Alex drew at this point] Then, if you apply the same force a second time to the same object, - well, it just moves faster.

He subsequently sketched and described the (sinusoidal and essentially infinitely long) wave as a series of pulses and described each pulse exerting a force like a kick or a hit on the dust particle. As before, the collision itself is consistent with an object-like description of the sound wave, while the description of repeated hits indicates that Alex is also able to think in event-like terms.

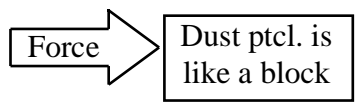

**Figure 3**

Alex's sketch of the sound wave exerting a force on the dust particle. Alex described the wave exerting a force on the dust particle (and later candle flame) only in the direction of wave propagation.

A more detailed discussion of Alex's thinking shows how he used an inappropriate set of resources to guide his reasoning about the physics. Alex stated, 'the frequency of the wave, a normal wave, shows us how many cycles per some period of time we have'. This interpretation of frequency focuses on the manner in which we describe events. But, as seen by his explicit use of forces acting on objects, Alex instead focused on object-like reasoning as he focused on the 'hits' on the dust particle.

The description that the sound wave exerts a force only in the direction of wave propagation shows that Alex thinks of the leading edge of the wave pushing everything in front of it away from the sound source, much like a surfer riding on an ocean wave. At a later point in the interview, he explained the surfer analogy by describing the motion of a ring on a string on which a pulse is propagating. The effect of a pulse on a small ring placed on the string was to push it along.[4] He used the term 'impulse' to describe the wavepulse and the effect of the wavepulse on its surroundings.

Alex: This impulse will hit the ring here, and ... [the impulse] should go and make [the ring] move forward, the same way it should be with a dust particle in the air.

When asked about the effect of changing the volume of the sound produced, he stated the following:

Alex: I guess I'm not thinking physics too much. [I'm thinking of a] stereo system at home, if you turn it up, you can feel the vibration from farther away from the speaker, so basically [the dust particle] should move, once again, it should move faster.
I: What effect did changing the volume have on the compression wave?
A: Increased the amplitude...
I: And that has the effect of the compression wave moving faster?
A: Not quite, it just hits the particle with more force. ... If you kick the thing, instead of kicking it faster, you're just kicking it harder. It's going to move faster.

Again, Alex described the effect of the wave exerting a force only in the direction of propagation to make the dust particle move forward. Note also his opening comment, that he is not 'thinking of physics' and is instead thinking of examples from his daily life. In subsequent conversation, Alex described that he was a sound studio engineer, and was thinking of the loud music 'hitting [his] chest'. Here, we see a clear example of the way resources formed in daily life may play a role in student reasoning in the classroom. Overall, we see how Alex can be thinking in terms of both events and objects, but the object-like description of waves predominates in his reasoning.

Of the six students who participated in our preliminary set of interviews, four gave similar descriptions of the effects of the sound wave on the dust particle. To document the richness of possible responses that students can give when answering questions on sound waves with which they are not familiar, we discuss the responses of the other two students. One student gave responses which were inconsistent, stating the correct answer (horizontal oscillation) while also stating that the dust particle would not move. Even with continued questioning, the student was unable to provide a clear response, showing that a profound confusion lay behind the student's correct responses. It is possible that this student would perform very well on examinations where the student is aware of the correct answers that the instructor is seeking, but still not have an actual understanding of the physics of sound waves. More specifically, without the ability to choose between conflicting models, the student is unable to know (outside the artificial constraints of an exam) when to apply the correct physics.

The other student interpreted the common sinusoidal graph used to describe sound waves (either pressure or displacement from equilibrium as a function of time or position) as a picture rather than a graph (Trowbridge & McDermott, 1980, 1981; Beichner, 1994) and seemed to use this misinterpretation to guide his reasoning. He described the motion of the dust particle as transverse and predicted that the candle flame would not move since it was unable move up or down, due to its attachment to the wick. This student misinterpreted the common sinusoidal graph of sound waves (where the *vertical* axis describes *horizontal* displacement from equilibrium as a function of position or of time), and used this misinterpretation to guide his understanding of the motion of the medium. He stated that the longitudinal compression of the sound wave would squeeze the dust particle to push it up or suck it back down (due to the 'vacuum' caused by a sort of rarefaction between longitudinal waves). The longitudinal wave would cause transverse motion in the dust particle. Note that this response contains many elements of correct physics (e.g. the rarefaction phase). The detailed physical explanation this student gave indicates the extent to which students will go to reconcile the physics with their interpretations (in this case, reading a graph as a picture) to make sense of the physics they learn.

In addition to the dust particle version of the question, these six students also answered the candle flame question shown in figure 1(b). Of the four students who gave responses like Alex's, above, three described the candle flame consistently: the sound wave hits the candle and knocks it sideways. The sound wave would either hold the candle there permanently or come at discrete moments in such a way that the candle would oscillate between the upright and pushed over (away from the speaker) state. The student who answered this question differently in the dust particle and candle case described the candlelight as energy which should not be affected by a sound wave. Of the other two students, one did not answer the question due to time limitations, and the other's response has already been described.

## Sound waves guiding the medium along a sinusoidal path

In a second set of interviews, more than twenty students were given the dust particle in multiple-choice format approximately two months after students had completed all instruction in waves. Some of these students had participated in early versions of the Tutorial materials described in more detail below, while others had completed traditional instruction on waves. Due to the interview format, it was possible to follow up on student responses and ask for clarification even on multiple choice questions.

One student, Kyle, described sound waves pushing dust particles in a sinusoidal path away from a loudspeaker. As he stated,

Kyle: The dust particle will move up and down and the dust particle will be pushed away from the speaker.

Interviewer: So the dust particle is going to move in a path away from the speaker (indicates a sinusoidal path with a hand motion in front of Kyle)

K: Yeah.

I: Why is that? If you could explain...

K: ... it will push the dust particle sideways. Since the dust particle is affected by the frequency, as long as the frequency is constant, it will move in a constant path.

Kyle was describing the dust particle moving along the path of the sinusoidal sine wave.(For similar responses in the context of physical optics, see Ambrose, Shaffer, Steinberg, & McDermott, 1999) (Linder, 1993) By 'constant path,' he was indicating movement along an unchanging sinusoid. Later, when asked the effect of a change in the frequency, Kyle explained that

> K: [The dust particle] will move faster because the frequency is higher which means that, since frequency is the one that affects the dust particle, if the frequency increases, the speed of the particle should increase, also.

Though on the surface this might indicate a correct response (a higher frequency will cause a faster average speed, since the particle is oscillating more rapidly), Kyle seemed to refer to the motion of the particle away from the speaker. Kyle often used the phrase 'the frequency affects the dust particle,' and when asked what he meant by the phrase, he stated

> K: Frequency produces a sound wave, and the sound wave, somehow it will... (He did not finish the statement, even with prodding from the interviewer).

In this description, Kyle is not using the term frequency to describe a property of the sound wave, rather, he states that the frequency causes the sound wave. Thus, the sound wave pushes the dust particle because 'the dust particle is affected by the frequency' and 'if the frequency increases, the speed of the dust particle should increase, also'. Kyle's explanation seems consistent with the explanation given by Alex above. Kyle is describing a wave whose motion and speed away from the speaker is determined by the properties of the wave that is pushing it forward.

In Kyle's explanation, the sound wave seems to guide the particle along the sinusoidal path. In his description of the particle's sinusoidal motion, he states that the dust particle must move along the path determined by the sinusoidal sound waves. This is an example of the reasoning resource, Guiding, described in other situations by diSessa (diSessa, 1993). Also, Kyle uses the idea of Working Harder (diSessa, 1993); stating that the greater the frequency, the greater the speed of the dust particle.

Like Alex, Kyle uses a mixture of reasoning resources that can be helpful in interpreting both events and objects. The 'up and down' motion of the dust particle, is indicative of his thinking about the former, while Working Harder is indicative of the latter.

## Summary of student responses

Other students gave similar explanations to Kyle's. In this second set of interviews, more than half the students spoke of a sound wave 'hitting' or 'pushing' objects out of its way. Rather than focus on what students were not doing (recognizing the rarefaction cycle of sound wave oscillations), we focus on what specifically was leading them to produce these answers. The students seem to think about sound waves using a common-sense interpretation based on sound waves exerting a one-directional force on the medium through which they travel. In summary, the students

- map object-like properties onto sound waves,
- treat them as solid and pushing through a medium, and
- do not correctly interpret the event-like properties that are more appropriate in this setting.

## ***Discussion of student performance, traditional instruction***

Based on interviews and student responses to written examination questions, we have found that lecture instruction alone (with the associated homework problems being discussed in a traditional recitation setting) had little effect on student understanding of the basic wave properties of sound propagation. Table 1 shows (unmatched) student responses from the beginning of spring, 1996, and end of fall, 1995, engineering university physics classes when answering a question like the one in figure 1. This early version of the question showed the loudspeaker enclosed in walls to form a tube. A non-trivial number of students (roughly 10% in both cases, listed within the 'other oscillation' category) sketched standing wave patterns (e.g. sinusoidal standing waves with nodes or antinodes at the end of the tube) to describe the motion of the dust particle. The tube walls were removed in later questions to remove this source of student confusion, but the result is an important one. Students seem to pick the familiar details or surface features of a problem to guide their reasoning in their responses. The tubes seemed to a activate a response based on common diagrams with which they were familiar, but this response showed the manner in which students can give responses which build on totally different reasoning resources than the ones relevant in a given situation.

Table 2 shows student explanations from the fall, 1997 semester to the dust particle question before instruction, after traditional instruction, and after additional modified instruction (described in more detail in the next section of this paper). We see that very few of the students enter our courses using the accepted physics description of sound wave propagation. Instead, they focus on the pushing or hitting description given above. Before instruction, half the students state that the sound wave pushes the dust particle away from the speaker. Some, like Alex, describe the dust particle moving in a straight-line path. Others, like Kyle, describe the dust particle moving along a sinusoidal path away from the speaker. The latter students seem to

| Time during semester: MM used: | Before all instruction S96 (%) | Post lecture F95 (%) |
|---|---|---|
| Longitudinal oscillation | 14 | 24 |
| Other oscillation | 17 | 22 |
| Pushed away linearly or sinusoidally | 45 | 40 |
| Other and blank | 24 | 14 |

**Table 1**

Comparison of student responses describing the motion of a dust particle due to a loudspeaker. Data are from Fall 1995 post instruction and Spring 1996 pre-instruction and are not matched (S96, N=104. F95, N=96).

| Time during semester: Explanation: | Before all instruction (%) | Post lecture (%) | Post lecture, post tutorial (%) |
|---|---|---|---|
| Longitudinal oscillation | 9 | 26 | 45 |
| Other oscillation | 23 | 22 | 18 |
| Pushed away linearly or sinusoidally | 50 | 39 | 11 |
| Other | 7 | 12 | 6 |
| Blank | 11 | 2 | 21 |

**Table 2**

Student performance on sound wave questions before, after traditional lecture, and after additional modified tutorial instruction. Data are matched (N=137 students). The large number of blank responses in the post-all instruction category is due to the number of students who did not complete the pretest on which the question was asked.

misinterpret the sinusoidal graph of displacement from equilibrium as a picture of the path of the particle. Their response only makes sense if the sinusoidally guided path is reasonable, which it is in the case of a particle being pushed by the sound wave. After traditional instruction, student responses to the sound wave questions are not notably improved. Though some change is seen in the number of students giving the correct response, the performance is still consistent with the performance shown in table 1. The improvement in scores due to the Tutorial instruction will be discussed after the Tutorial itself has been discussed in more detail.

## Curriculum Development To Address Student Reasoning

In this section, we describe how the curriculum we developed was designed to help students build on their previous appropriate knowledge. We then give evaluation results to describe the strengths and weaknesses of the material. We explicitly describe how the model of student learning played a role in curriculum development in order to help those instructors who wish to understand and apply our more general theoretical approach rather than simply use the specific materials we created. By making our goals and research-basis explicit, we hope to help instructors who might modify these materials to better match their own setting and also those who interact with students in their classroom and encounter responses not specifically described in this paper.

### Description of Tutorial materials

The sound Tutorial builds on our research results, outlined above, which are consistent with how students approach other areas of wave physics (Wittmann et al., 1999; Wittmann, 2001). The focus on object-like wave properties (that exert unidirectional forces on things in their way, for example) is obviously problematic, and we attempt to provide students with a new description of the physics that is based on skills and appropriate knowledge that they already bring to the classroom, but do not apply to this situation. Rather than describe sound waves that exert forces as if they were able to kick or hit things in their way, we focus on the movement of similar events from one point to another. (For a recommendation for this instruction in the case of waves on a string, see Arons, 1990.) In the interviews, we found that students were able to use event-oriented reasoning in some contexts. In the Tutorial, we try to strengthen their skills in this area. The materials described below encompassed one hour of instruction in addition to the traditional instruction students had already received on the basic physics of sound. Lecture instruction for the fall 1997 class described in the remainder of this paper included such advanced topics such as the Doppler effect and was independent of the tutorial. The same amount of time was spent on instruction in sound during the fall 1997 semester as in the spring, 1996, and fall, 1995, semesters.

Though the majority of the research was carried out in the context of the dust particle question, the Tutorial begins with the motion of a candle flame placed in front of a loudspeaker. Data from a pretest from fall 1997 (see table 3) show that students have generally the same difficulties describing the motion of a candle flame that they have describing the motion of a dust particle. Only matched data are presented in the table.[5] These results are consistent with the interview results, showing that the candle flame context provides a fair opportunity for students to develop their reasoning about sound waves.

| Object Whose Motion is Being Described explanation: | Dust Particle (%) | Candle Flame (%) |
|---|---|---|
| Longitudinal oscillation | 22 | 39 |
| Other oscillation | 26 | 3 |
| Pushed away linearly or sinusoidally | 38 | 21 |
| Other | 11 | 17 |
| Blank | 3 | 20 |

**Table 3**
    Performance on student pretest, comparing descriptions of dust particle and candle flame motion. Students answered the two questions at the same time (F97, data are matched, N=215). The high number of blank responses on the candle flame question is due to lack of time during the pretest, which was followed by an mid-term examination.

    In the Tutorial, students begin by predicting and then viewing a digitised video of the motion of a single candle due to a sound wave coming from a large loudspeaker (see figure 4[6]). As in the interviews, many students in the classroom predict that the candle will be 'hit' and pushed over, but never pulled back to the loudspeaker. These students are confronted with the video showing motion to both sides of the equilibrium position of the candle (i.e. including motion toward the loudspeaker). They must account for the entire motion of the candle and resolve discrepancies between their prediction and the observation. In addition, we ask that they explicitly apply their predictions to the context of the dust particle that was part of the pretest.
    In the next section of the Tutorial, students are given data that shows the position of the left edge of the candle flame at different times. The data points have been taken beforehand using the program VideoPoint (Luetzelschwab & Laws, 2001) and are presented to the students in a data table in the Tutorial. (Due to time limitations during the Tutorial, students are not asked to take the data themselves.) In the Tutorial, students must observe the connection between the data points and the cross-mark on the left edge of the candle on the screen. Students are asked to graph the data points on a provided graph. From the graph, they then find the period of the

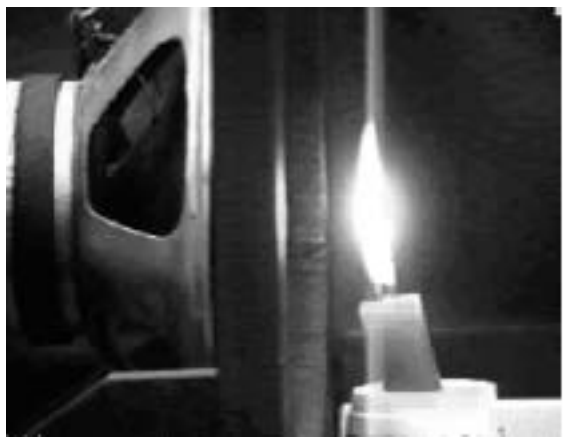

**Figure 4**
    Candle flame and loudspeaker in a digitised video frame. Students must compare the candle's side-to-side oscillation with their predictions of its movement.

candle's oscillation. In the activity, students go from a description of a single candle (which represents the motion of the medium) to describing the frequency of the sound wave. Thus, they are given the opportunity to connect observations, mathematical descriptions, and physical properties that they have discussed in class and used in their homework.

Students are then presented with a photograph of two candles sitting in front of a loudspeaker. They are asked to describe the motion of both candles and to sketch separate displacement vs. time graphs for each candle. To answer the question, students must generalize from their previous description of a single candle's motion to think about any possible changes between the motion of the first and the second candle. Students must use the idea of a propagating wave with a finite speed to develop the idea of a phase difference between the motion of the two candles. The two candles experience identical events due to identical causes, but at different times due to the wave propagation speed.

This idea is developed through a Gedankenexperiment where students are asked to think of more and more candles placed at different locations along a path away from the speaker. They are asked to sketch the displacement from equilibrium of each candle at a specific instant in time. From this activity, they can find the wavelength of the sound wave. Again, students are given the opportunity to connect their mathematical knowledge from class with reasoning based on simple ideas of wave physics. Furthermore, these activities help reinforce the idea that a propagating wave causes identical events to happen (at different times, due to finite wave speed) throughout the system. The idea of a wave as a propagating event or occurrence is thereby emphasized.

The sound Tutorial provides an opportunity for students to develop basic concepts of wave physics in an understandable and familiar setting. In a previous Tutorial, students had discussed wave propagation and the motion of waves through a medium.(This Tutorial was designed based on research into student reasoning described in Wittmann, 1998) In that Tutorial, they discussed visible waves travelling along taut strings. In this Tutorial they discuss the propagation of unseen waves. This provides an opportunity for students to make connections between the visible and invisible, a topic of great importance as they continue their studies with such topics as fields, light waves, and temperature flow. In another Tutorial (Steinberg et al., 1997), students are introduced to the mathematical description of wave propagation for wavepulses (Arons, 1990). In the sound Tutorial, they apply this idea to sinusoidal wavetrains. Students have an opportunity to bring the familiar mathematics into an otherwise unseen, abstract realm of everyday experience. They integrate the many representations to give a picture of the physics consistent with their observations.

More importantly, students are encouraged to use a more appropriate set of reasoning tools than the ones with which they entered the course. We have observed that they can think in event-oriented terms, but often think about sound waves in object-like terms instead. The sound Tutorial gives them a chance to think of waves causing events at two different locations with a phase lag in between. This focus on occurrences at two different locations forces students to use a different set of tools with which to think about waves. The use of appropriate observations and representations that they must interpret brings the students to question different elements of the their system and build an event-oriented description that best fits the tasks.

The Tutorial provides an excellent context in which students can interpret waves from a non-object-like point of view. Rather than focusing on waves 'hitting' the candle, students must make connections between the loudspeaker which creates the wave and the candle which moves due to the wave. Furthermore, they must look at the behaviour of candles at different locations

                    

going through the same motions. Thus, the wave is interpreted as a propagating event or occurrence, rather than a moving unidirectional force on the system.

Once students have completed the Tutorial, they have an additional brief homework assignment designed to incorporate and revisit many questions from the Tutorial in a new context. Students again build connections between the mathematical, graphical, and physical interpretations of the situation while applying event-oriented reasoning to the physics of sound waves.

### Student understanding of sound waves

Results from previous semesters' courses and student performance before and after traditional instruction are shown in tables 1 and 2. The third column in table 2 indicates student performance after Tutorial instruction. Students were given the post-test question without announcement two months after all instruction on waves. Note that the large number of students who left the question blank is most likely due to a lack of class time in which to answer all the questions on the (ungraded) post-test.

After the Tutorial, a much larger number of students are able to describe the correct motion of the oscillating dust particle. Note that their responses were categorized as 'other oscillations' when they did not specify the oscillation direction or stated the movement was in the vertical direction. Though students describing vertical oscillation might have difficulties with the common representations used to describe sinusoidal sine waves, their reasoning is of a different nature than that given by students describing a unidirectional force acting on the dust particle.

Though the improvement in student performance is encouraging, the gain is not as large as is desired. Nearly 2/3 of those students answering the question describe a longitudinal oscillation, but over 10% still describe the sound wave exerting a unidirectional pushing force.[7] Still, the results are encouraging when viewed from the point of view that students were not expecting the post-test, had not studied the material recently, and knew they were not being graded on the material. In the next section, we describe related questions that probed students' thinking on the issues in a context very different from the candle flame or dust particle questions. These also showed that students' abilities to reason about sound waves had improved.

### Discussion of inconsistent student reasoning in the classroom

To investigate student consistency on sound wave questions, four questions on the fall, 1997, pre- and post-test addressed sound-wave physics. So far, only one has been discussed. The three additional questions were all of the same nature as the candle-flame questions in that students had to describe the effects of sound waves on the medium the waves pass through or the effects of wave properties (e.g. volume, frequency) on the speed of wave propagation. An example of such a question is shown in figure 5, where students were asked to compare wave speed of the sounds created when two different people yelled at each other. In this question, answers that stated that different pitches or amplitudes were indicative of different wave speeds (e.g. consistent with the descriptions given by Alex and other students) were judged to be consistent with an object-like description of sound waves. An accepted correct answer would have been to state that the waves travelled at the same speed.

Figure 6 shows the number of students that answered using an object-like model of waves and the number that used a wave physics model in a bubble plot (where the area of the bubble describes the number of students). Since students often gave multiple explanations for a single

> Michael and Laura are standing 100 m apart and yell
> "Yo!" at each other at exactly the same instant.
> Michael yells louder than Laura, and the pitch
> (frequency) of his voice is lower. No wind is blowing.
>
> Will Laura hear Michael first, Michael hear Laura
> first, or will they hear each other at the same time?
> Explain how you arrived at your answer.

**Figure 5**

Example sound wave diagnostic test question. Student responses were evaluated according to whether they interpreted the initial conditions (volume and frequency) according to object-like or wave-like properties.

response, it was possible for the total number of student responses to be greater than four when they used both object-like and wave physics reasoning. Only matched data are included in our comparison. 141 students answered at least three of the four questions both before and after instruction. Note that each bubble represents how many students gave a specific number of object-like and a specific number of wave physics responses. Before instruction, only one student gave 4 event-like and 0 object-like responses.

The plot (figure 6) shows that students are neither consistent nor coherent in their answers either before and after instruction. Most students begin the semester answering predominately with object-like reasoning but give some event-like responses. After instruction, we find that more students answer the four sound-wave questions using both object-like and wave-like reasoning and few give only wave-like responses.

To describe the results less broadly requires an analysis in terms of the different resources that students apply to the different questions. Sometimes the students are triggered to use resources involving the forces exerted on the dust particle, and they often give object-like responses (involving unidirectional forces, for example). At other times, these same students think of wave behaviour and the manner in which waves pass through the medium, and they therefore give event-based wave physics responses. Note that in the previously described interviews, given in a slightly different setting, students also did not answer such questions consistently.

As is clear from the discussion of student responses in a variety of settings, students are able to think in multiple ways about a single situation. This result can have a serious effect when an instructor is working in a classroom. This observation is consistent with other contexts of wave physics (Wittmann et al., 1999). The story of curriculum development based on insight into the resources that students use can help instructors in their teaching outside of the specific context of the curriculum discussed in this paper.

Recognizing that students bring multiple tools to the classroom in terms of the resources and sets of resources they use can help us make sense of their behaviour in a variety of settings. The context dependence of student responses plays a central role in how an instructor approaches students in the classroom. An instructor listening to students cannot assume that their correct model of the physics in one setting implies that they are consistent in their reasoning. Their additional models must be investigated (e.g. through an appropriate question that might activate a different viewpoint and explanation of the physics).

                    

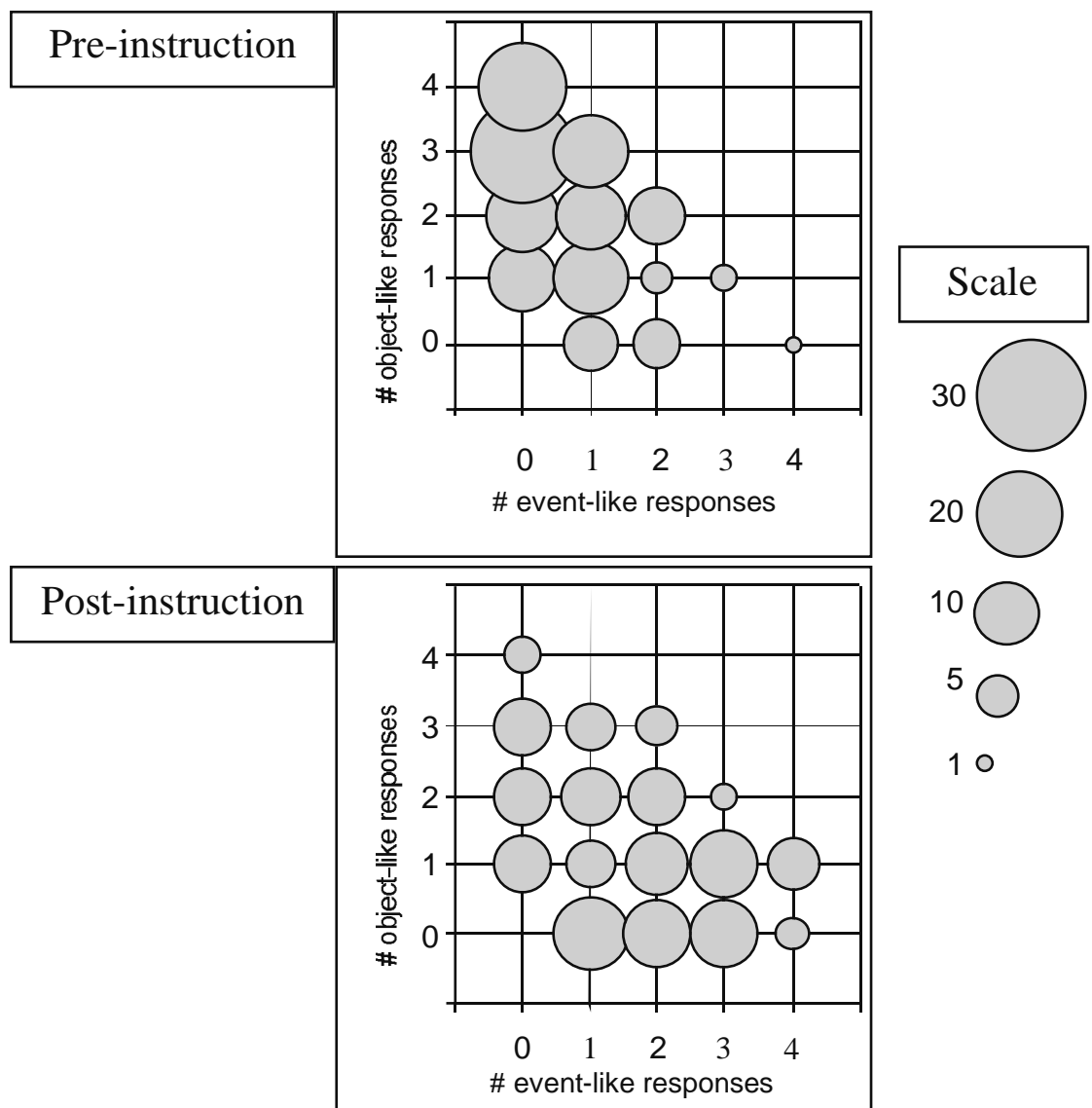

**Figure 6**

Bubble plot of student responses (matched, n=141) to four pre-and post-instruction sound wave test questions. Size of bubble represents number of students giving a specific number of event- and object-like responses.

For researchers, such situations provide an interesting area in which to study the development of student sense-making and consistency-checking mechanisms. Possible paths of inquiry involve an analysis of the epistemologies students have, and how these play a role in their sense-making of the material. Do students believe that physics is a coherent body of knowledge? Do they see the inconsistency in their thinking when giving contradictory answers to nearly identical situations? Work by Hammer (Hammer, 2000) and Elby (Elby, 2001) explores these issues in more detail.

Another path of inquiry would be to model a class's performance on a test (or question) designed to distinguish between possible different approaches to the physics. Work by Bao and Redish (Bao & Redish, in preparation) illustrates how to analyse group performance on such questions. Though this approach washes away indications of which contexts specifically cause problems for the students and also which students are most mixed in their reasoning, it does indicate to an instructor the likelihood of encountering mixed models among the students in the classroom. Few well-designed tests and well-formulated analysis mechanisms exist at the moment for instructors to use to get a quick view of their class at the beginning of an instructional unit.

## Conclusion

In this paper, we describe how we have developed curriculum materials by focusing on the fine-grained reasoning elements that students often apply to inappropriate settings. We explicitly describe a set of productive reasoning elements that students use in analysing how sound waves act on light objects. By illustrating this process, we hope to inform not just the teachers who might use these materials in the classroom, but also the researchers who are engaged in similar projects.

For teachers, we hope to have presented an approach that will allow them to act flexibly in their classroom to meet the needs of their specific students. Instructors responding to students in the classroom might recognize the nature of student difficulties and adapt flexibly by emphasizing the productive resources that help students learn the physics. In our setting, the Tutorial has been successful. In other settings, additional questions focusing on stadium waves, dominoes falling over, or similar intermediate systems might be helpful. Further research is required to evaluate the effectiveness of these curricula, and to discern which examples are best at building up the appropriate productive resources for understanding wave propagation.

The instructional materials we developed focus on helping students apply a set of reasoning resources they were not using to full potential when describing sound wave physics. Rather than focusing on object-like reasoning which focuses on sound waves exerting unidirectional forces on the medium through which they travel, students are given the opportunity to develop ideas based on similar events occurring at different locations and different times due to a propagating wave. By promoting this language in a situation where we have found that students are able to use event-like reasoning but tend to activate primarily object-like reasoning, we help students evaluate when to use which set of resources appropriately. In addition, students also discuss and describe a variety of representational and conceptual elements and how they are related. All of these can lead to the observed improvement in student performance on a variety of sound wave questions in pre- and post-instruction tests.

The instructional materials have met with measurable but qualified success. More students are able to answer a variety of sound wave questions successfully after completing Tutorial instruction, but a significant percentage of the class still uses object-like reasoning after instruction. In addition, students do not think about the physics consistently after the additional Tutorial instruction. These results raise issues both for instruction, in terms of which teaching methods can best help students move toward consistent use of the appropriate physics, and for research, in terms of questions about the role of epistemology, probabilistic models of student reasoning, and appropriate models of learning.

## Acknowledgements

The work described in this paper was carried out with the contributions of the entire Physics Education Research Group at the University of Maryland. In particular, Mel Sabella filmed the video used in this Tutorial while at the summer workshop of Workshop Physics,

RealTime Physics, and Interactive Lecture Demonstrations at Dickinson College, PA, in 1996. He also developed the original Tutorial materials. Discussions with Andy Elby led to a greatly improved paper. We would also like to thank the National Science Foundation and the Department of Education Fund for Improvement of Post-Secondary Education (FIPSE) for their financial support in this research.

## Endnotes

---

[1] 'The wave' is common in large sporting stadiums in the US. One stands when one's neighbour stands (with a bit of a lag), one sits when the neighbour sits. When all spectators participate, a circularly travelling wave of standing and sitting people moves around the stadium. This nearest-neighbour interaction causing a chain of successive events is an excellent example of wave motion.

[2] In both questions, we assumed audible frequency sound waves, between 10 and roughly 5,000 Hz. Assuming a speed of sound of roughly 340 m/s, this gives a range of wavelengths between 7 cm and 34 m. All of these wavelengths are much greater than the size of either the dust particle or the candle flame (roughly 1/2 to 1 cm wide). The shortest wavelengths occur at a frequency that is already outside of the common frequencies heard on a daily basis in speech or in music. (The highest of these are usually around 2000 Hz, giving a wavelength of 17 cm.

[3] Note that all names used in this paper are aliases chosen by the students. These students were volunteers who were usually performing at the top of their class. We use dashes to indicate interruptions, ellipses (…) to indicate deletions of unimportant text, and square brackets to insert explanatory elaborations. Pauses in speech are usually not indicated.

[4] He gave a partially correct answer for a ring that is on a string; the wavepulse will make the ring move in some longitudinal fashion that depends on the cosine of the angle the string makes with its undisturbed position as the wavepulse passes by. For small angles this effect will be small. Alex gave an incomplete description, not speaking of rarefactions where the effect would be opposite.

[5] Note that the high number of blank responses on the candle flame response may be due to the candle flame question coming in the later half of the pretest. In one of the two classes that took the pretest, it was given in a 75 minute class on the same day as a one hour mid-term examination. Students were not given much time for the pretest, and many students simply did not attempt the majority of it.

[6] The video of the oscillating candle is available on the WWW at URL http://perlnet.umephy.maine.edu/perl/materials/candle.mov.

[7] Though attendance is not taken in tutorials, it is usually quite high in comparison to standard-instruction discussion sections. We do not know the number of students taking the post test who participated in the sound wave tutorial.